\documentclass{amsproc}
\usepackage{amssymb}

\newtheorem{theorem}{Theorem}[section]

\newtheorem{lemma}[theorem]{Lemma}

\theoremstyle{remark}
\newtheorem*{remark}{Remark}

\numberwithin{equation}{section}

\newcommand{\bbR}{{\mathbb{R}}}
\newcommand{\calH}{{\mathcal H}}

\newcommand{\lb}{\label}
\newcommand{\f}{\frac}
\newcommand{\veps}{\varepsilon}
\newcommand{\bi}{\bibitem}

\DeclareMathOperator{\Exp}{Exp}

\begin{document}

\title{A Feynman-Kac Formula for Unbounded Semigroups}

\author[B.~Simon]{Barry Simon}
\address{Division of Physics, Mathematics, and Astronomy, 
253-37 \\ California Institute of Technology \\ Pasadena, CA~91125, USA.}
\email{bsimon@caltech.edu}
\thanks{This material is based upon work 
supported by the National Science Foundation under Grant 
No.~DMS-9707661. The Government has certain rights in this material.}
\thanks{To appear in Proceedings of the International Conference 
on Infinite Dimensional (Stochastic) Analysis and Quantum Physics, 
Leipzig, 1999}


\begin{abstract} We prove a Feynman-Kac formula for Schr\"odinger 
operators with potentials $V(x)$ that obey (for all $\veps >0$) 
\[
V(x) \geq -\veps |x|^2 - C_\veps.
\]
Even though $e^{-tH}$ is an unbounded operator, any $\varphi, \psi 
\in L^2$ with compact support lie in $D(e^{-tH})$ and $\langle \varphi, 
e^{-tH}\psi\rangle$ is given by a Feynman-Kac formula.
\end{abstract}
\maketitle

\section{Introduction} \lb{s1}

One of the most useful tools in the study of Schr\"odinger operators, 
both conceptually and analytically, is the Feynman-Kac formula. All the 
standard proofs, (see, e.g., \cite{SFI}) assume the Schr\"odinger operator $H$ 
is bounded below, so the Schr\"odinger semigroup $e^{-tH}$ is 
bounded. This means, for example, that Stark Hamiltonians are not included.

But the restriction to semibounded $H$ is psychological, not real. We deal 
with unbounded $H$'s all the time, so why not unbounded $e^{-tH}$? Once 
one considers the possibility, the technical problems are mild, and it is 
the purpose of this note to show that.

The form of the Feynman-Kac formula we will discuss is in terms of the 
Brownian bridge (Theorem~6.6 of \cite{SFI}). Once one has this, it is easy 
to extend to the various alternate forms of the Feynman-Kac formula.

The $\nu$-dimensional Brownian bridge consists of $\nu$ jointly Gaussian processes, 
$\{\alpha_i (t)\}^\nu_{i=1; 0\leq t\leq 1}$ with covariance
\begin{gather*}
E(\alpha_i (t)\alpha_j(s)) = \delta_{ij} \min(t,s) 
[1-\max(t,s)] \\
E(\alpha_i (t)) = 0.
\end{gather*}
If $b$ is Brownian motion, then $\boldsymbol{\alpha}(s) = \boldsymbol{b}(s)
-s\boldsymbol{b}(1)$ is an explicit realization of the Brownian bridge.

For any real function $V$ on $\bbR^\nu$ and $t>0$, define (the expectation 
may be infinite):
\begin{equation} \lb{1.1}
Q(x,y; V,t) = E\left(\exp \left( - \int_0^t V\left( 
\left(1-\f{s}{t}\right) x + \f{s}{t}\, y + \sqrt t\, 
\alpha \left(\f{s}{t}\right)\right)\, ds \right)\right). 
\end{equation}
Throughout this paper, let
\[
H_0 = -\tfrac{1}{2}\Delta
\]
on $L^2 (\bbR^\nu)$, so
\begin{equation} \lb{1.2}
e^{-tH_0} (x,y) = (2\pi t)^{-\nu/2} \exp \left( -\f{|x-y|^2}{2t}\right). 
\end{equation}
The Feynman-Kac formula I'll start with --- one of many in \cite{SFI} --- is

\begin{theorem} \label{t1.1} Suppose $V$ is a continuous function on 
$\bbR^\nu$ which is bounded from below. Let $H=H_0 +V$. Then for any 
$t>0$ and $\varphi,\psi \in L^2 (\bbR^\nu)$:
\begin{equation} \lb{1.3}
\langle \varphi, e^{-tH}\psi\rangle = \int \overline{\varphi(x)}\, 
\psi(y)\, e^{-tH_0} (x,y) Q(x,y;V,t).
\end{equation}
\end{theorem}

In this paper, we will consider potentials $V(x)$ for which for any $\veps 
>0$, there is $C_\veps$ so that
\begin{equation} \lb{1.4}
V(x) \geq -\veps |x|^2 - C_\veps. 
\end{equation}

It is known (see \cite{RS2}, Theorem~X.38) that for such $V$, $H=H_0 + 
V$ is essentially self-adjoint on $C^\infty_0 (\bbR^\nu)$, so we can use 
the functional calculus to define $e^{-tH}$ which might be unbounded. Our 
main goal here is to prove:

\begin{theorem} \lb{t1.2} Suppose $V$ is a continuous function which obeys 
\eqref{1.4}. Then for all $x,y\in\bbR^\nu$, $t>0$, \eqref{1.1} is finite. 
Let $\varphi,\psi\in L^2 (\bbR^\nu)$ have compact support. Then for all 
$t>0$, $\varphi,\psi \in D(e^{-tH})$ and \eqref{1.3} holds.
\end{theorem}

\textsc{Remarks}. 1. It isn't necessary to suppose that $\varphi, 
\psi$ have compact support. Our proof shows that it suffices that 
$e^{\veps x^2}\psi, e^{\veps x^2}\varphi\in L^2$ for some $\veps >0$. In 
particular, $\varphi,\psi$ can be Gaussian.

\smallskip
2. Using standard techniques \cite{AHK},\cite{CZ},\cite{SFI}, one can 
extend the proof to handle $V=V_1 + V_2$ where $V_1$ obeys \eqref{1.4} 
but is otherwise in $L^1_{\text{\rm{loc}}}$ and $V_2$ is in the Kato 
class, $K_\nu$.

\smallskip
3. If one only has $V(x) \geq -C_1 - C_2 x^2$ for a fixed $C_2$, our proof 
shows that the Feynman-Kac formula holds for $t$ sufficiently small. It may 
not hold if $t$ is large since it will happen if $V(x) = -x^2$ that 
$E(\exp(-\int_0^t V(\alpha (s))\, ds))$ will diverge if $t$ is large.

\medskip

As for applications of Theorem~\ref{t1.2}, one should be able to obtain 
various regularity theorems as in \cite{SSS}. Moreover, for $H=-\Delta + 
\boldsymbol{F} \cdot \boldsymbol{x}$, one can compute $e^{-tH}(x,y)$ 
explicitly and so obtain another proof of the explicit formula of Avron 
and Herbst \cite{AH}.

\medskip
\noindent{\bf Dedication.} Sergio Albeverio has been a master of using and extending the 
notion of path integrals. It is a pleasure to dedicate this to him on the 
occasion of his 60th birthday.

\medskip

\section{A Priori Bounds on Path Integrals} \lb{s2}

Our goal in this section is to prove
\begin{theorem} \lb{t2.1} Let $V$ obey \eqref{1.4} and let $Q$ be given 
by \eqref{1.1}. Then, for each $t>0$ and $\delta >0$, we have that
\[
Q(x,y; V,t) \leq D\exp(\delta x^2 + \delta y^2),
\]
where $D$ depends only on $t,\delta$ and the constants $\{C_\veps\}$.
\end{theorem}

\begin{lemma} \lb{L2.1} Let $X$ be a Gaussian random variable. Suppose 
$\veps\Exp(X^2)< \f12$. Then $E(\exp(\veps X^2))<\infty$ \rm{(}and is 
bounded by a function of $\veps\Exp(X^2)$ alone\rm{)}.
\end{lemma}

\begin{proof} A direct calculation. Alternately, we can normalize $X$ so
$\Exp(X^2) =1$. Then $E(\exp (\veps X^2))=(2\pi)^{-1/2} \int \exp ((\veps 
-\f12)x^2)\, dx < \infty$. 
\end{proof}

\begin{proof}[Proof of Theorem~\ref{t2.1}] Note that if $0<\theta<1$, and 
$x,y,\alpha\in \bbR^\nu$, then
\begin{align*}
|\theta x + (1 -\theta)y + \alpha|^2 &\leq 2 |\theta x + (1-\theta)y|^2 
+ 2|\alpha|^2 \\
&\leq 2(x^2 + y^2 + |\alpha|^2).
\end{align*}
Thus, by \eqref{1.4},
\begin{equation} \lb{2.1}
Q(x,y; V,t) \leq E\left(\exp \left( C_\veps t + 2 \veps t(x^2 + y^2) + 
2\veps \int_0^1 t^2 \alpha (s)^2\, ds \right)\right).
\end{equation}

By Jensen's inequality,
\begin{equation} \lb{2.2}
E\left( \exp\left( 2 \int_0^1 \veps t^2 \alpha(s)^2\, ds \right)\right) \leq 
\int_0^1 E(\exp(2\veps t^2 \alpha(s)^2)\, ds).
\end{equation}
Since $E(\alpha(s)^2)$ is maximized at $s=\f12$ when it is $\f14$, we see 
that
\[
\text{RHS of }\eqref{2.2} \leq E(\exp (2\veps t^2 \alpha(\tfrac12)^2))
\]
is finite if $\veps t^2 <1$, so we can pick $\veps = \delta_0/t^2$ with 
$\delta_0 <1$ and find (using the explicit value of $E(\exp(X^2)$) in 
that case
\[
Q(x,y; V,t) \leq \sqrt 2\, (1-\delta_0)^{-1/2} \exp(C_\veps t + 2\delta_0 
(x^2 + y^2) /t),
\]
which proves Theorem~\ref{t2.1}.
\end{proof}

\medskip

\section{A Convergence Lemma} \lb{s3}

In this section, we will prove:

\begin{theorem} \lb{t3.1} Let $A_n, A$ be self-adjoint operators on a 
Hilbert space $\calH$ so that $A_n \to A$ in strong resolvent sense. Let 
$f$ be a continuous function on $\bbR$ and $\psi\in\calH$ with $\psi\in 
D(f(A_n))$ for all $n$. Then
\begin{enumerate}
\item[(i)] If $\sup_n \|f(A_n)\psi\| <\infty$, then $\psi\in D(f(A))$.
\item[(ii)] If $\sup_n \| f(A_n)^2 \psi\|<\infty$, then $f(A_n)\psi 
\to f(A) \psi$.
\end{enumerate}
\end{theorem}

\begin{remark} Let $\calH =L^2 (0,1)$, $\psi(x)\equiv 1$, $A_n =$ 
multiplication by $n^{1/2}$ times the characteristic function $[0, 1/n]$, 
and $A\equiv 0$. Then $A_n \to A$ in strong resolvent sense and $\sup_n 
\|A_n \psi\|<\infty$, but $A_n\psi$ does not converge to $A\psi$ so one 
needs more than $\sup_n \|f(A_n)\psi\|<\infty$ to conclude that $f(A_n)\psi 
\to f(A)\psi$. The square is overkill. We need only $\sup_n \|F(f(A_n))\psi\| 
<\infty$ for some function $F:\bbR\to\bbR$ with $\lim_{|x|\to\infty} |F(x)| /x
=\infty$.
\end{remark}

\smallskip

\begin{proof} Suppose that $\sup_n \|f(A_n)\psi\|<\infty$. Let 
\[
f_m(x) = \begin{cases} m &\text{if } f(x) \geq m \\
f(x) & \text{if } |f(x)| \leq m \\
-m &\text{if } f(x) \leq -m.
\end{cases}
\]
Then (\cite{RS1}, Theorem~VIII.20) for each fixed $m$, $f_m (A_n)\to 
f_m(A)$ strongly. It follows that 
\begin{align*}
\|f_m (A) \psi\| &= \lim_n \|f_m(A_n)\psi \| \\
&\leq \sup_n \| f_m (A_n)\psi\| \leq \sup_n \| f(A_n)\psi\|.
\end{align*}
Thus, $\sup_m \|f_m(A)\psi\|<\infty$, which implies that $\psi\in D(f(A))$.

Now suppose $\sup_n \|f(A_n)^2 \psi\| <\infty$. Then
\[
\| (f(A_n) - f_m (A_n)) \psi \| \leq \f{1}{m}\, \|f(A_n)^2 \psi\|.
\]
Thus $f_m (A_n)\psi \to f(A_n)\psi$ uniformly in $n$ which, given that 
$f_m (A_n)\psi\to f_m(A)\psi$, implies that $f(A_n)\psi\to f(A)\psi$. 
\end{proof}

\medskip

\section{Putting It Together} \lb{s4}

We are now ready to prove Theorem~\ref{t1.2}. Let $V$ be continuous and 
obey \eqref{1.4}. Let $V_n (x) = \max(V(x), -n)$. Then $V_n$ is bounded 
from below, so Theorem~\ref{t1.1} applies, and so \eqref{1.3} holds. 
Let $\varphi\in L^2$ with compact support. By Theorem~\ref{t2.1}, we have
\[
\sup_n \| \exp(-tH_n)\varphi\| <\infty
\]
for each $t$ positive.

By the essential self-adjointness of $H$ on $C^\infty_0 (\bbR^\nu)$ and 
$(V_n - V)\eta \to 0$ for any $\eta\in C^\infty_0$, we see that $H_n$ 
converges to $H$ in strong resolvent sense. Hence setting $A_n = H_n$, 
$A=H$, $f(x)=e^{-tx}$, and $\psi=\varphi$, we can use Theorem~\ref{t3.1} 
to see that $\varphi\in D(\exp(-tH))$ and $\|[\exp(-tH_n) - \exp(-tH)]
\varphi\|\to 0$. Thus as $n\to\infty$, the left-hand side of the Feynman-Kac 
formula converges. By the a priori bound in Theorem~\ref{t2.1} and the 
dominated convergence theorem, the right-hand side converges. So 
Theorem~\ref{t1.2} is proven.

\medskip


\end{document}